\documentclass[letterpaper,aps,prl,twocolumn,showpacs,superscriptaddress,amsmath,amssymb]{revtex4}

\usepackage{graphicx}
\usepackage[latin1]{inputenc}
\usepackage{dcolumn}
\usepackage{hyperref}

\begin{document}

\title{Generation and Characterization of Multimode Quantum Frequency Combs}

\author{Olivier Pinel}
\affiliation{Laboratoire Kastler Brossel, Université Pierre et Marie Curie--Paris 6, ENS, CNRS; 4 place Jussieu, 75252 Paris, France}
\author{Pu Jian}
\affiliation{Laboratoire Kastler Brossel, Université Pierre et Marie Curie--Paris 6, ENS, CNRS; 4 place Jussieu, 75252 Paris, France}
\author{Renné Medeiros de Ara\'ujo}
\affiliation{Laboratoire Kastler Brossel, Université Pierre et Marie Curie--Paris 6, ENS, CNRS; 4 place Jussieu, 75252 Paris, France}
\author{Jinxia Feng}
\affiliation{Laboratoire Kastler Brossel, Université Pierre et Marie Curie--Paris 6, ENS, CNRS; 4 place Jussieu, 75252 Paris, France}
\affiliation{State Key Laboratory of Quantum Optics and Quantum Optics devices, Institute of Opto-Electronics, Shanxi University, Taiyuan 030006, P. R. China}
\author{Benoît Chalopin}
\affiliation{Laboratoire Kastler Brossel, Université Pierre et Marie Curie--Paris 6, ENS, CNRS; 4 place Jussieu, 75252 Paris, France}
\affiliation{Max Planck Institute for the Science of Light, Universität Erlangen-Nürnberg, IOIP, Staudtstrasse 7/B2, 91058 Erlangen, Germany}
\author{Claude Fabre}
\email{claude.fabre@upmc.fr}
\affiliation{Laboratoire Kastler Brossel, Université Pierre et Marie Curie--Paris 6, ENS, CNRS; 4 place Jussieu, 75252 Paris, France}
\author{Nicolas Treps}
\affiliation{Laboratoire Kastler Brossel, Université Pierre et Marie Curie--Paris 6, ENS, CNRS; 4 place Jussieu, 75252 Paris, France}

\date{\today}

\pacs{42.50.Dv, 42.50.Lc, 42.65.Yj}

\begin{abstract}
Multimode nonclassical states of light are an essential resource in quantum computation with continuous variables, for example in cluster state computation. We report in this paper the first experimental evidence of a multimode non-classical frequency comb in a femtosecond synchronously pumped optical parametric oscillator.  In addition to a global reduction of its quantum intensity fluctuations, the system features quantum correlations between different parts of its frequency spectrum. This allows us to show that the frequency comb is composed of several uncorrelated eigenmodes having specific spectral shapes, two of them at least being squeezed, and to characterize their spectral shapes.

\end{abstract}

\maketitle

Optical frequency combs are perfect tools for high precision metrological applications \cite{Udem2002, Holzwarth2000}. The extension of their extraordinary properties to the quantum domain may lead to significant progress in different areas of quantum physics, in particular in quantum metrology and parameter estimation \cite{Lamine2008, Pinel2010}, but also in quantum computation with continuous variables \cite{Lloyd1999,Menicucci2006}. Indeed, one of the main challenges of experimentally implementing quantum computers in the continuous variable regime, for example in cluster state computation \cite{Zhang2006, Menicucci2006}, is the generation of highly multimode non-classical states of light, and the scalability of this generation. As the difficulty of linearly mixing distinct squeezed light sources \cite{Yukawa2008, Obrien2009} increases as the number of modes increases, it can be more interesting to use instead a single highly multimode source which directly produces non-classical resources shared between many modes within a same beam. In this perspective, optical frequency combs,  which span over thousands of different frequency modes, are a very promising system for scalable generation of spectral/temporal multimode quantum states. We report in this paper the first experimental evidence of multimode non-classical frequency comb generated by an Optical Parametric Oscillator (OPO) in the femtosecond regime, which opens the way to the generation of these highly multimode states. 

Multimode non-classical light has been already experimentally generated with spatial multimode beams produced by OPOs \cite{Janousek2009, Chalopin2010}, and very recently, with the longitudinal modes of an OPO \cite{Menicucci2008, Pysher2011}. In the domain of temporal modes, single mode squeezing of short pulses has been observed in various experiments starting from \cite{Slusher1987} in the nanosecond regime. Non-classical states of single femtosecond pulses are the subject of many recent studies (for example \cite{Wenger2005}). Multimode squeezed solitons have been  generated in an optical fiber \cite{Spalter1998b}. Single mode quantum noise reduction in picosecond frequency combs has already been achieved with a Synchronously Pumped Optical Parametric Oscillator (SPOPO) \cite{Shelby1992}, which is an OPO pumped by a train of ultrashort pulses that are synchronized with the pulses making round trips inside the optical cavity. 

It has recently been shown \cite{deValcarcel2006, Patera2009} that such SPOPOs generate squeezed frequency combs which are multimode. We give in this paper the experimental confirmation of these predictions. The different squeezed modes are actually frequency combs having different spectral profiles, or equivalently in the time domain, trains of pulses having different temporal profiles.

In our experiment, the frequency comb is produced by a singly-resonant SPOPO pumped with pulses in the femtosecond range, which are described in the frequency domain by a superposition of at least $10^5$ longitudinal modes of frequencies $\omega_n^\mathrm{p}$ located around the carrier frequency $2\omega_0$ and equally spaced by a repetition rate $\omega_\mathrm{r}$: $\omega_n^\mathrm{p}=2\omega_0+n\omega_\mathrm{r}$. This huge number of pump modes leads to a great complexity of the parametric down conversion process taking place in the intracavity nonlinear crystal. Indeed, each pumping frequency $\omega_n^\mathrm{p}$ is coupled through phase matched parametric interaction to many pairs of cavity-resonant frequencies $\omega_\ell^\mathrm{s}$ and $\omega_{n-\ell}^\mathrm{s}$, where $\omega_\ell^\mathrm{s}=\omega_0+\ell\omega_\mathrm{r}$, since they satisfy $\omega_n^\mathrm{p}=\omega_\ell^\mathrm{s} + \omega_{n-\ell}^\mathrm{s}$. However, it has been demonstrated \cite{deValcarcel2006} that this interaction can be described by a substantially reduced number of modes which are the eigenmodes of the nonlinear interaction, named supermodes. These are well-defined coherent superposition of longitudinal modes characterized by their spectral amplitude and phase profiles. The mode profiles of the supermodes in the frequency domain, or in the time domain, are predicted to be close to Hermite-Gaussian functions \cite{Patera2009}.

\begin{figure}[htbp]
\includegraphics[width=8cm]{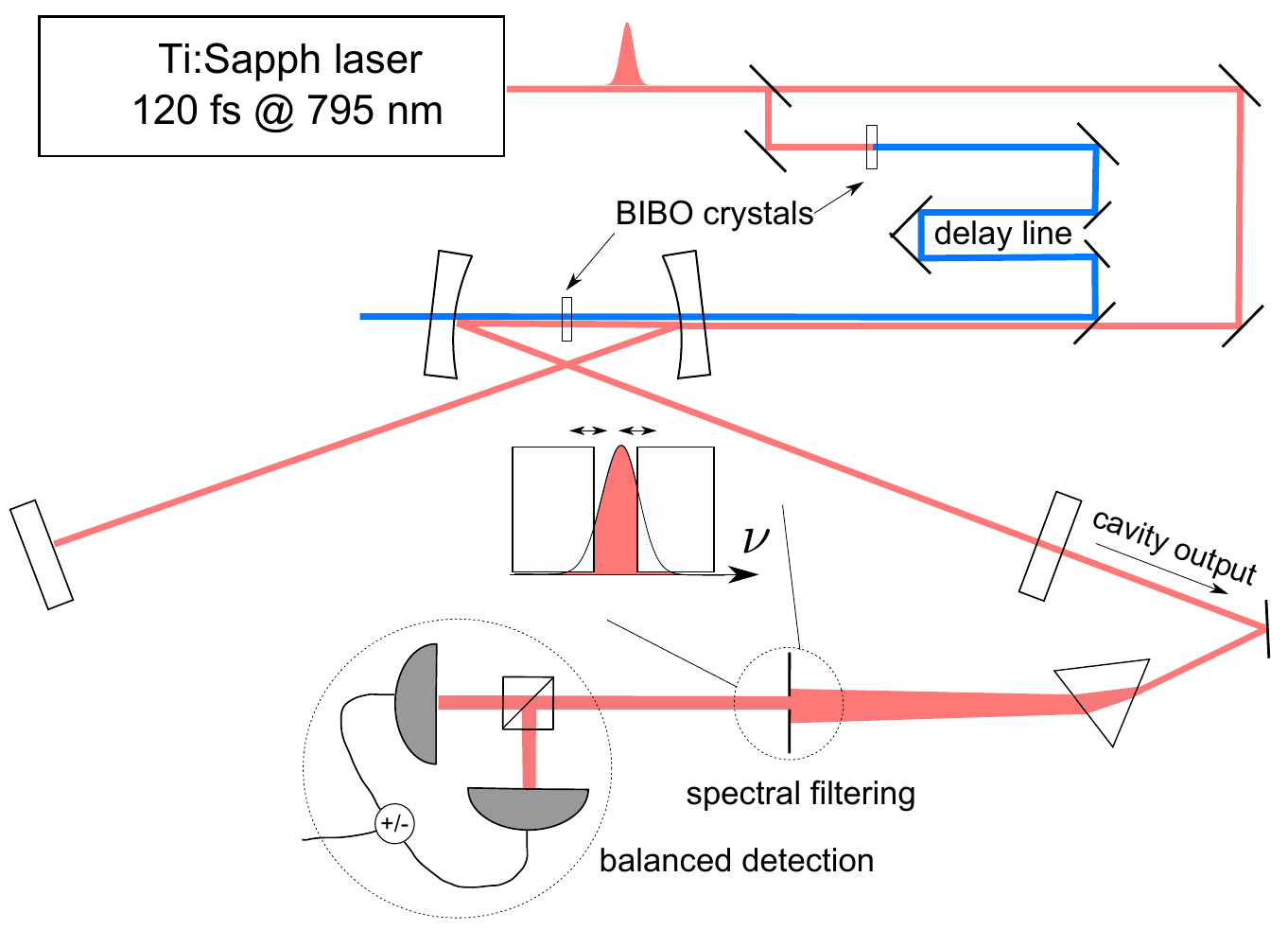}
\caption{\label{fig:setup}Schematic of the experimental setup. The OPO is synchronously pumped by a frequency comb centered at 397~nm. The delay line ensures the temporal overlap between the pulses from the pump and seed at cavity input. At the cavity output, the deamplified frequency comb centered at 795~nm is spectrally filtered. A balanced detection is used to measure the intensity noise.}
\end{figure}

\begin{figure}[htbp]
\includegraphics[width=8cm]{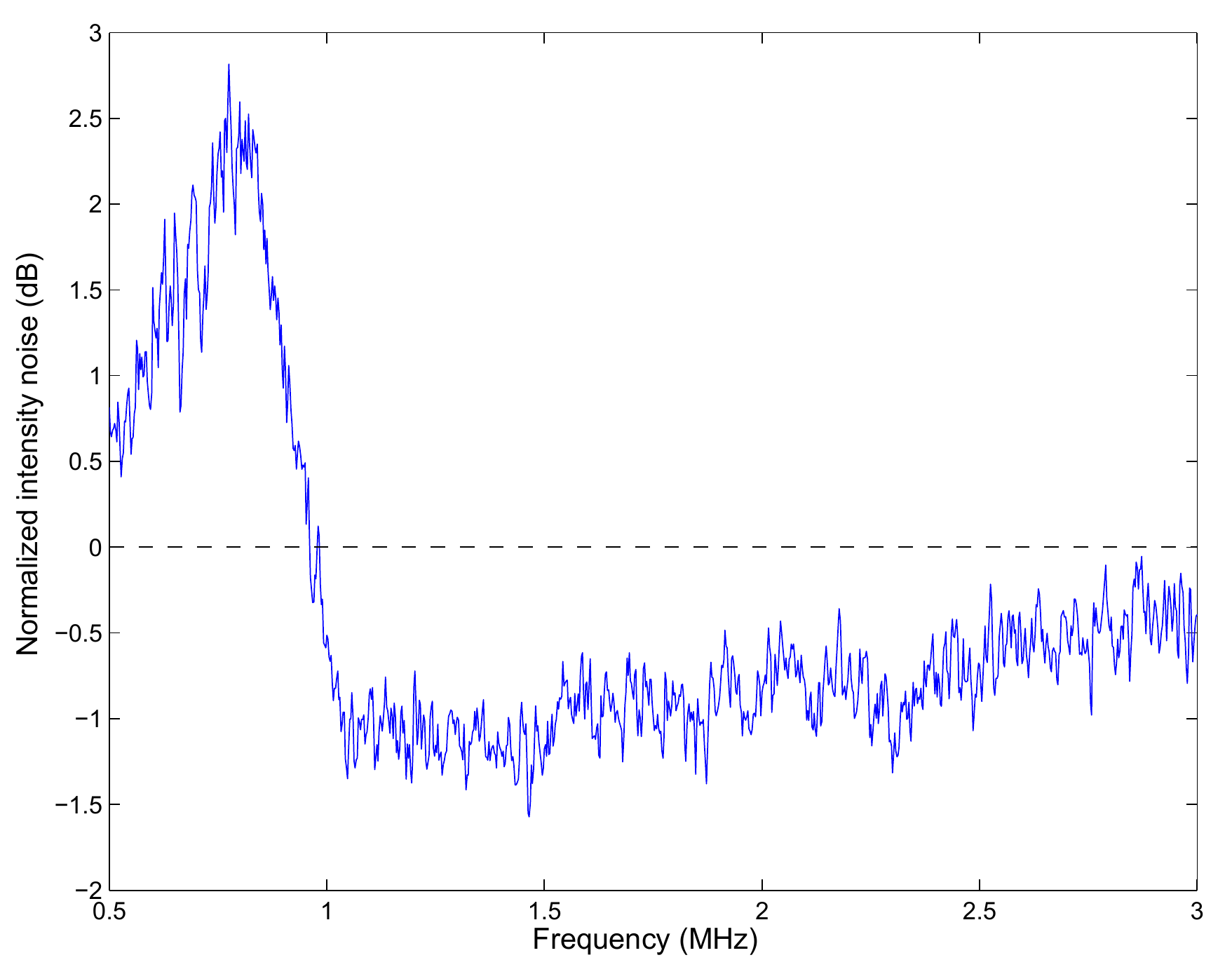}
\caption{\label{fig:squeezing}Spectral intensity noise (in dB) as a function of the analysis frequency on the spectrum analyzer. Dashed line: standard quantum limit. The spike below 1~MHz is due to the relaxation oscillation noise of the laser. These data are not corrected from the electronic dark noise of the photodiodes. Resolution bandwidth: 30~kHz; video bandwidth: 10~Hz.}
\end{figure}

The experimental setup is shown in Fig.~\ref{fig:setup}. The SPOPO is seeded by 120~fs pulses at 795~nm with a repetition rate of 76~MHz produced by a Ti:Sapphire mode-locked laser (where the carrier-envelop phase is not stabilized) pumped by a Nd:YVO$_4$ laser at 532~nm. Its second harmonic at 397~nm is used to pump the 350~$\mu$m long intracavity BIBO crystal \cite{Hellwig1998}. The SPOPO cavity length is locked using a Pound-Drever-Hall (PDH) lock. Above a threshold of typically $\sim 50$~mW the SPOPO generates a single signal-idler frequency comb having the same mean frequency as the Ti:Sapph laser (degenerate configuration). Below threshold, we observe phase-sensitive amplification of the seed. We lock the relative phase between the seed and the pump in the deamplification regime using another PDH lock. The state of the output light is evaluated using a two-port balanced detector (with quantum efficiency above $90\%$ and 30~dB noise extinction between the two detectors): the sum of the photocurrent fluctuations represents the intensity fluctuations $\Delta n^2$ of the output beam, and the difference represents the standard-quantum-limited fluctuations of a beam of same power $\Delta n_\mathrm{shot}^2 = \langle n \rangle$. We measure a normalized intensity noise up to ${\Delta n^2}/{\Delta n_\mathrm{shot}^2}=0.76\pm0.02$ at $\sim1.5$~MHz, corresponding to $1.2\pm0.1$~dB of noise reduction on the amplitude quadrature (see Fig.~\ref{fig:squeezing}). This experimentally demonstrates for the first time the non-classicality of the field generated by a femtosecond SPOPO below threshold. The low amount of squeezing may be explained by the fact that the seed is not optimized in the present status of the experiment: it does not coincide with the SPOPO supermode with highest achievable squeezing  (later referred as the first supermode). Indeed the spectrum of this supermode, which depends on the spectrum of the pump and the length of the crystal, is theoretically 8.3 times broader than the spectrum of the seed. The field we measure with the balanced detection therefore corresponds to a superposition of different supermodes, which results in a higher intensity noise than in the first supermode. In addition, as the cavity frequency bandwidth at half maximum is 2.5~MHz, more squeezing should be observed at a lower noise frequency, which is presently not possible because of the presence of excess technical noise (due to the relaxation oscillation of the laser) at low frequencies of analysis.

\begin{figure}[htbp]
\includegraphics[width=8cm]{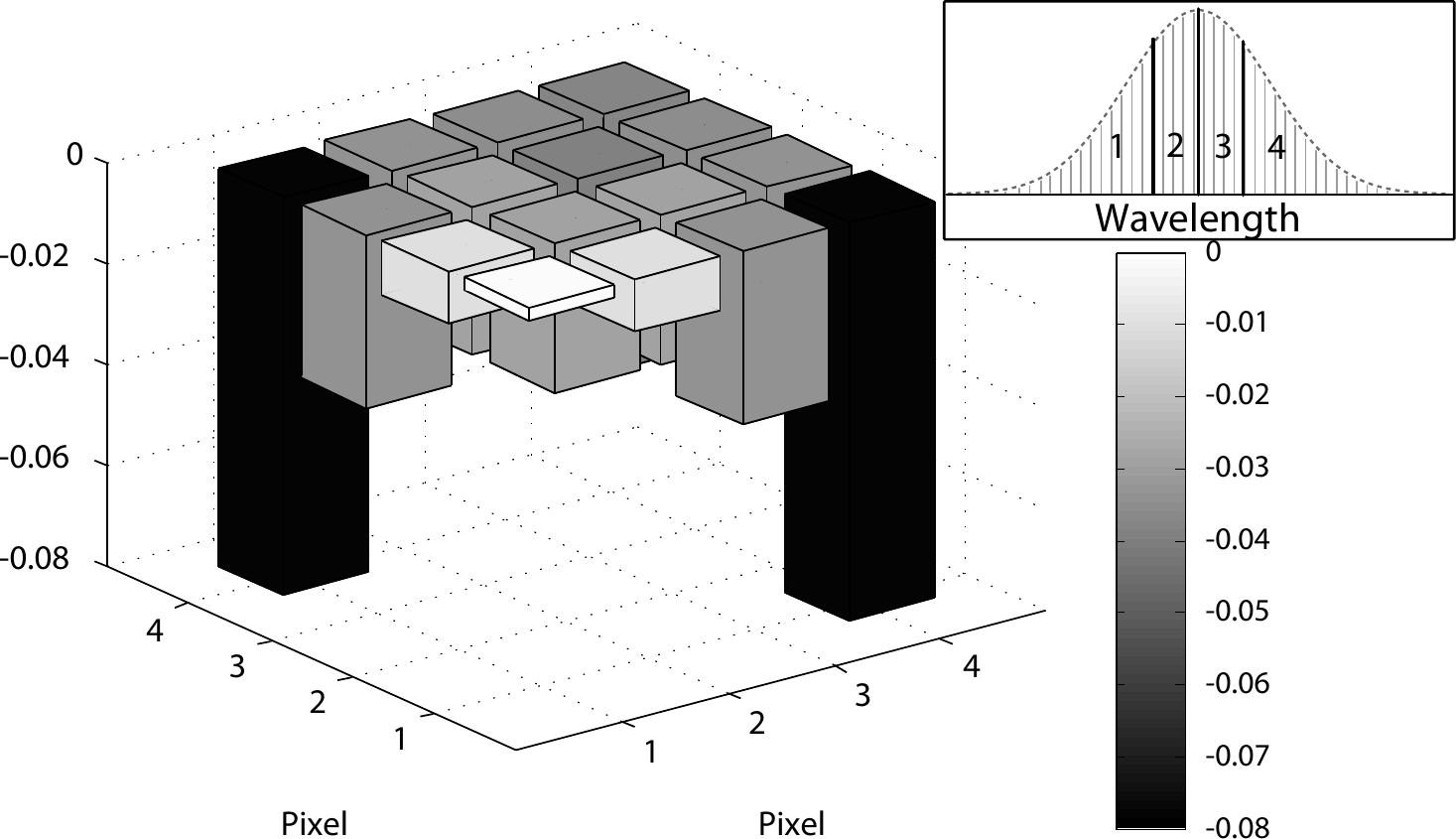}
\caption{\label{fig:correlation} Mean normalized photon number correlation matrix $C(i,j)$. Inset: schematic of the frequency pixels compared to the spectrum.}
\end{figure}

In order to investigate the multimode nature of the output beam, we have studied the distribution of the quantum intensity fluctuations of the frequency comb over its optical spectrum, in a way similar to \cite{Spalter1998b, Opatrny2002}. To this aim the frequency components of the output beam were separated using two prisms (only one prism is shown on Fig.~\ref{fig:setup} for sake of simplicity). A slit of variable position and width allowed us to record intensity fluctuations of different parts of the spectrum. The spectral resolution of the filter was 1.8~nm.  More precisely, we divided the frequency comb spectrum into four consecutive frequency pixels of equal mean intensities and symmetric with respect to the central frequency (see inset of Fig.~\ref{fig:correlation}). We recorded the intensity noise and the shot noise level for different frequency intervals: in addition to the measurements of pixels $1,\ldots,4$ and of the whole spectrum, we also measured combinations of two pixels $\{1,2\},\{2,3\},\{3,4\}$ and of three pixels $\{1,2,3\},\{2,3,4\}$, leading to a total of 10 frequency intervals. For each frequency interval, we measured successively the intensity noise and the shot noise during 5~seconds, corresponding to 1\,000 data points for each measurement. The SPOPO remained locked during all the measurement process. The mean values were used to calculate the normalized intensity noise for each frequency interval.

For each frequency interval formed by the sum of pixels $\{i_1,\ldots,i_m\}$, where $i_1$ and $i_m \geq i_1$ are integers between 1 and 4, the detected intensity fluctuations are 
\begin{equation}
\Delta\big(\sum_{i=i_1,\ldots,i_m}n_i\big)^2 = \sum_{i,j=i_1,\ldots,i_m}\textrm{cov}(n_i,n_j)
\end{equation}
where  $\textrm{cov}(n_i,n_j)=\langle{n_in_j}\rangle-\langle{n_i}\rangle\langle{n_j}\rangle$ is the photon number correlation function between the different pixels when $n_i \neq n_j$ and the variance when $n_i = n_j$ . From these measurements, the intensity correlation matrix 
\begin{equation}
C(i,j) = \frac{\textrm{cov}(n_i,n_j)}{\sqrt{\Delta n_i^2 \Delta n_j^2}} - \delta_{ij}\frac{\Delta n_{i,\mathrm{shot}}^2}{\Delta n_i^2}
\end{equation}
can be reconstructed. The mean matrix is plotted in Fig.~\ref{fig:correlation}. As the four pixels have the same mean power, it is simple to show that, in the case of a single mode state, all these correlation functions should be equal to -0.035. The presence of one diagonal element (1,1) with less local squeezing and of one off-diagonal element (1,4) with more anti-correlation than the other coefficients suggests that the output of the SPOPO is indeed a multimode non-classical field. However, the uncertainty on the off-diagonal elements does not allow us to unambiguously state the the generated state is multimode.

In order to demonstrate the multimode character, we performed another analysis of the output of the SPOPO which consists in calculating the eigenmodes involved in the present experiment and to show if these modes are indeed excited. To this purpose we reconstructed the covariance matrix in the basis of the four frequency pixels:
\begin{equation}
V_{x_i,x_j}=\frac{1}{2}\langle{\delta x_i\delta x_j+\delta x_j\delta x_i}\rangle
\end{equation}
where $x_i=a_i+a_i^\dagger$ is the amplitude quadrature, and $\delta x = x - \langle{x}\rangle$. The phase quadrature information cannot be recovered from our technique as we measure only intensity noises. We make therefore the simple assumption that the different frequency components of the output field have the same phase, and can all be taken as real. Using the derivation discussed in \cite{Opatrny2002}, one finds $\textrm{cov}(n_i,n_j)\approx \langle{x_i}\rangle\langle{x_j}\rangle V_{x_i,x_j}$. The mean photon number in one zone is taken as $\langle{n_i}\rangle\approx\langle{x_i}\rangle^2$, leading to the following expression for the elements of the covariance matrix:
\begin{equation}
V_{x_i,x_j}=\frac{\textrm{cov}(n_i,n_j)}{\sqrt{\Delta n_{i,\mathrm{shot}}^2 \Delta n_{j,\mathrm{shot}}^2}}
\end{equation}

\begin{figure}[htbp]
\includegraphics[width=8cm]{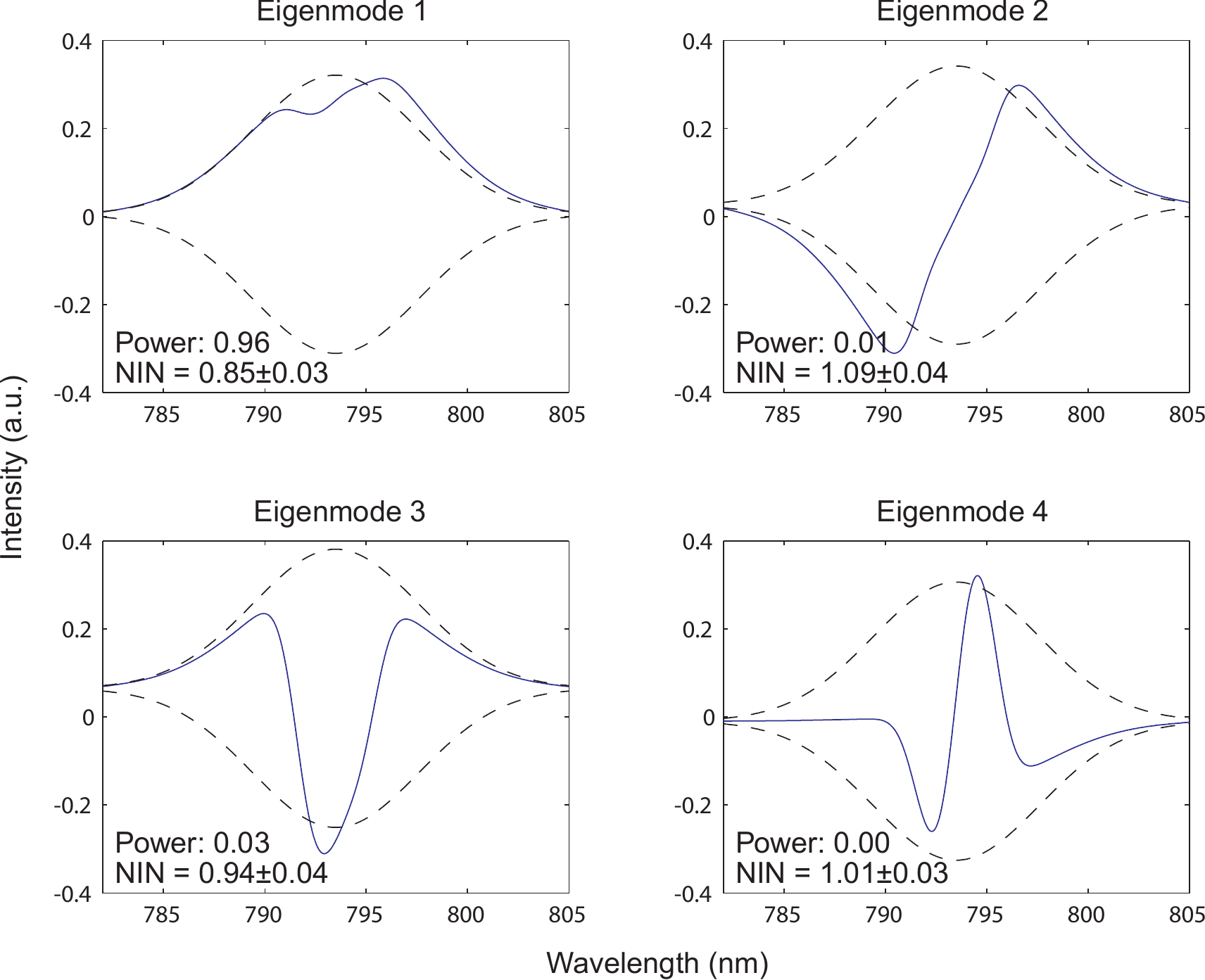}
\caption{\label{fig:eigenmodes}Wavelength profile of the eigenmodes $S_\ell$. The pixel modes are smoothed by taking into account the 1.8nm spectral resolution of the bandpass filters. Solid lines: eigenmodes; dashed line: mean field mode. Power: percentage in optical power of the output beam; NIN: normalized intensity noise.}
\end{figure}

The diagonalization of the covariance matrix $V$ allows us to find that the light generated by the SPOPO is made of a set of four uncorrelated modes (the eigenvectors of $V$), with given amplitude noise variances (the eigenvalues of $V$). We find that among these four eigenmodes $S_\ell$, two have eigenvalues smaller than the standard quantum limit~1, and one has an eigenvalue greater than~1, as shown in Fig.~\ref{fig:eigenmodes}. The uncertainty in the measurement is small enough to conclude that the first and the third mode are amplitude squeezed and that the second has excess noise compared to vacuum, but not to conclude whether the fourth mode is excited. This shows that the output of the SPOPO is described by at least 3 independent modes, two of them at least being in a squeezed state; the fourth mode may be considered as vacuum. The uncertainty on the intensity noise of the eigenmodes was estimated in the following way: we generated 10\,000 photon number covariance matrices with elements randomly picked among our experimental data points. After rotation in the $S_\ell$ basis, the off-diagonal elements of these matrices fluctuate around $0\pm0.03$, showing that the modes $S_l$ are effectively eigenmodes of the randomly generated fields. The diagonal elements give the normalized intensity noise of the eigenmodes, and their spread give the spread of the normalized intensity noise.

Furthermore, the eigenmode $S_1$, within which lies most of the power of the light beam, has approximately the same amount of squeezing than the mean field. The spectrum of $S_1$ is slightly broader than the mean field mode, which is consistent with the theoretical prediction \cite{Patera2009} for the first supermode in our experimental conditions. We also observe that the profile of $S_2$ resembles a Hermite-Gaussian mode of order 1, and shows excess noise; this is also consistent with the theory, which predicts that the second supermode of the SPOPO is a first order Hermite-Gaussian mode squeezed in phase, and therefore anti-squeezed in amplitude. In the same way, the profile of eigenmodes $S_3$ appears like a Hermite-Gaussian mode of order 2, and $S_3$ is squeezed, as predicted by its even order. Finally, one notices a slight asymmetry in these modes, which may be related to group-velocity dispersion effects.
 
The procedure that we have described demonstrates that at least three orthogonal modes are necessary to describe the output field of the SPOPO, two at least being in squeezed states. However, a complete description of the field generated by the SPOPO may involve more modes: separating the spectrum in more than four pixels may reveal intensity correlations which would require more than three modes to be described. Moreover, this derivation does not take into account the possibility of correlations with the phase fluctuations $p_i$. A complete characterization of the quantum state of the output of the SPOPO requires to have access to $V_{p_i,p_j}$ and $V_{x_i,p_j}$, which requires to perform a homodyne detection with appropriately shaped local oscillator pulses \cite{Cundiff2010}.

In conclusion, we have experimentally demonstrated that femtosecond SPOPOs generate below threshold multimode quantum frequency combs that can be precisely characterized and analyzed in terms of squeezed uncorrelated eigenmodes of experimentally determined shapes. Indeed, for Gaussian states, multimode entanglement and multimode squeezing are equivalent given that one can choose the measurement basis \cite{Braunstein2005}. The low level of noise reduction and of quantum correlation makes it so far a proof-of-principle experiment. Noise reduction, and therefore possible entanglement, can be certainly drastically increased to much higher levels by working at lower noise frequencies with a less noisy pump laser and seeding it with a mode shaped beam close to the first supermode. In addition, it can be shown that the exact multimode quantum state of the generated light can be adjusted by controlling the pump pulse shape and duration. This experiment therefore opens the way to the production of quantum frequency combs that are tailored to fit the requirements of its numerous applications in quantum information processing and quantum metrology.

\begin{acknowledgments}
We acknowledge the financial support of the Future and Emerging Technologies (FET) programme within the Seventh Framework Programme for Research of the European Commission, under the FET-Open grant agreement HIDEAS, number FP7-ICT-221906; and of the ANR project QUALITIME.
\end{acknowledgments}

\end{document}